\documentclass[%
reprint,
superscriptaddress,
 amsmath,amssymb,
 aps,
]{revtex4-1}

\usepackage{graphicx}
\usepackage{dcolumn}
\usepackage{bm}

\usepackage[flushleft]{threeparttable} 

\usepackage{dsfont}
\usepackage{color,psfrag}
\usepackage{float}
\usepackage{soul}
\usepackage{xcolor} 
\definecolor{purple}{rgb}{0.5,0.0,0.5}
\definecolor{mypink1}{RGB}{219, 48, 122}
\definecolor{mypink2}{cmyk}{0, 0.7808, 0.4429, 0.1412}
\definecolor{mygray}{gray}{0.6}

\newcommand\here[1]{\fcolorbox{red}{red}{\rule{0pt}{6pt}\rule{6pt}{0pt}}\quad}

\usepackage{ifpdf}
    \ifpdf
\usepackage{tikz}
\usetikzlibrary{arrows,chains,matrix,positioning,scopes, fit, calc}
\usetikzlibrary{cd}
\usepackage{chemfig}
\fi
\usepackage{hyperref}

\hypersetup{
	colorlinks,
	linkcolor=blue,
	citecolor=blue, 
	filecolor=black,
	urlcolor=blue}

\begin{document}

\preprint{APS/123-QED}


\title{On the forbidden graphene's ZO (out-of-plane optic) phononic band-analog vibrational modes in fullerenes}

\author{Jes\'us N. Pedroza-Montero}
\affiliation{Programa de Doctorado en Nanociencias y Nanotecnolog\'ias, CINVESTAV, Av. Instituto Polit\'ecnico Nacional 2508, M\'exico}
\author{Ignacio L. Garz\'on}
\affiliation{Instituto de F\'isica, Universidad Nacional Aut\'onoma de M\'exico,
Apartado Postal 20-364, 01000 CDMX, M\'exico}
\author{Huziel E. Sauceda}
\email{sauceda@tu-berlin.de}
\email{Part of this work was done at Fritz-Haber-Institut der Max-Planck-Gesellschaft, 14195 Berlin, Germany} 
\affiliation{Machine Learning Group, Technische Universit\"at Berlin, 10587 Berlin, Germany
}
\affiliation{
 BASLEARN, BASF-TU joint Lab, Technische Universit\"at Berlin, 10587 Berlin, Germany
}

\date{\today}
\begin{abstract}
The study of nanostructures' vibrational properties is at the core of nanoscience research, they are known to represent a fingerprint of the system as well as to hint the underlying nature of chemical bonds. In this work we focus on addressing how does the vibrational density of states (VDOS) of the carbon fullerene family ($C_n:~n=20\to720$ atoms) evolves from the molecular to the bulk material (graphene) behavior using density functional theory. We found that the fullerene's VDOS smoothly converges to the graphene characteristic shape-line with the only noticeable discrepancy in the frequency range of the out-of-plane optic (ZO) phonon band in graphene. From a comparison of both systems we obtain as main results that: 1)The pentagonal faces in the fullerenes impede the existence of the analog of the high frequency graphene's ZO phonons, 2)which in the context of phonons this could be interpreted as a compression (by 43\%) of the ZO phonon band by decreasing its maximum allowed radial-optic vibration frequency. 3)As a result, the deviation of fullerene's VDOS relative to graphene should result on important thermodynamical implications. The obtained insights can be extrapolated to other structures containing pentagonal rings such as nanostructure or as pentagonal defects in graphene. 
\end{abstract}
\pacs{Valid PACS appear here}
\maketitle

\section{Introduction}
Since the discovery of the C$_{60}$ fullerene in 1985,~\cite{C60Buckminster1985} a cascade of theoretical and experimental studies on the physics and chemistry of novel carbon nanostructures emerged. As a consequence of the very interesting properties shown by carbon based nanomaterials, such as mechanical, electronic, optical, and chemical ones, a great number of fundamental investigations and technological applications have been developed since.\cite{nanoapp1,nanoapp2,nanoapp3,nanoapp4,nanoapp5,nanoapp6}
Despite the plethora of research done on carbon nanomaterials like fullerenes, nanotubes, nanoflakes, etc.,~\cite{ScalingLawVDW,VILLAGOMEZ_C20atAuSurf2020,Chen2020,Raimondo2020,Lima2020,Pochkaeva2020,Wang2020,Sha2020,AuNP-C60_2020} still some physicochemical properties have not been fully investigated. These include their vibrational properties and how they are related to their size and morphology, being the fullerene family a particular case.\\

During the last couple of decades, the vibrational properties of metal nanostructures have gained a lot of attention. \cite{Carnalla,Posada1996,Maioli_NanoLett2018} 
This is not only due to their relevance in the design of nanodevices,~\cite{Juve2010} but also because of a notorious development of  vibrational spectroscopies, opening the opportunity to directly compare experimental measurements and theoretical calculations.~\cite{Sauceda_VDOS_2015a}
For example, exploiting the symmetry breaking in supported metal nanoparticles (NPs), Carles' group developed a technique to directly measure the vibrational density of states (VDOS).~\cite{Bayle2014a,Bayle2014b}
Such a breakthrough~\cite{CarlesSR} lead to the understanding of intricate experimental results using well established theoretical results,~\cite{Cuenya2012,Sauceda2012,Sauceda_Na139_2013b,Lei2013,Sauceda_VDOS_2015a,Sauceda2015b} which link the VDOS as a NP vibrational fingerprint that correlates to the morphology of the system.~\cite{Posada1996,Sauceda_VDOS_2015a}
Contrasting these results, organic macro-molecules' spectroscopy relies only on Raman and IR spectroscopy to experimentally study them.
Hence, given the nature of these experimental measurements, we have access to only a selection of vibrational modes and not to their full VDOS.

A fundamental question in nanoscience is how the transition from molecular (physical and chemical) properties evolve to bulk behavior as a function of the system size.
In this regard, several results have been published such as the birth of the localized surface plasmon resonance in Au nanoclusters~\cite{Malola_Au-plasmon_ACSnano2013}, the evolution of thermodynamical properties on metallic nanoparticles~\cite{Fultz_PhysRevLett1997,Kara_JCTN2004,KARA_SurfSci2005,Cuenya2012}, the formation of surface plasmons in sodium nanoclusters~\cite{Na_optical_size2020}, as well as the convergence of molecular vibrational spectra to a bulk-like phonon density of states in metal NPs.~\cite{Sauceda_Au-NP_2013a}
In the particular case of metal NPs, a smooth transition of the VDOS from the nanoscale to the bulk has been reported, study that is missing in the case of carbon nanostructures.

In this work, we present a theoretical investigation of the vibrational properties of fullerenes and its dependence with size (20$\to$720 atoms) at the density functional theory (DFT) theory level. 
In particular, we analyse the evolution of the fullerene VDOS toward the bulk material (graphene) regime.
%
%
Analysis that leads to an interesting and counterintuitive difference between the vibrational modes in fullerenes and the phonon bands in graphene: a severe restriction on the out-of-plane (radial) optical band (ZO-modes) in fullerenes in the neighbourhood of the $\Gamma$ symmetric point.
These results show that the presence of the pentagonal faces in the fullerene family hinders the smooth convergence of their VDOS to the bulk graphene one.
Furthermore, theses findings hint that pentagonal impurities in graphene can have far more important implications on its vibrational properties than expected.
%

\begin{figure}[ht]
\includegraphics[width=0.95\columnwidth]{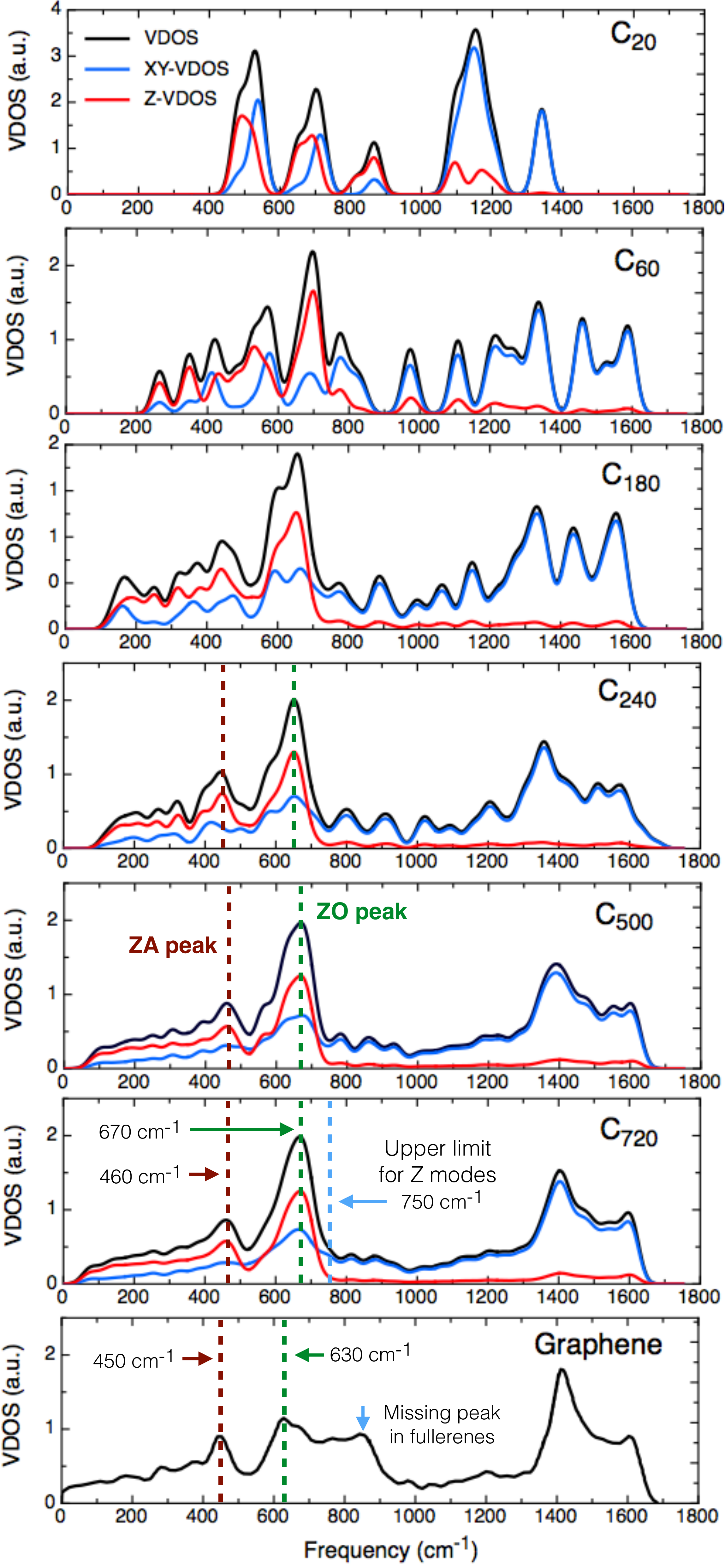}
\caption{VDOS (black curve) of fullerenes. The red and blue curves represent the radial and tangential contributions to the total VDOS, respectively. The VDOS was constructed using a Gaussian broadening with a width of 20 cm$^{-1}$. Bulk graphene VDOS (bottom panel) was taken from Ref.~\cite{Graphene2007}}
\label{fig:Fig_VDOSsize}
\end{figure}

\section{Results and discussion}
\subsection{Vibrational density of states} 
\subsubsection{The C$_{20}$ case}
Figure~\ref{fig:Fig_VDOSsize} shows the VDOS of fullerenes from 20 to 720 atoms (diameters from 0.4 to 2.6 nm) with $I_h$ symmetry. 
As a consequence of the finite size of the fullerenes, the frequency spectrum is discrete and it has a finite \textit{acoustic gap} $\nu_{\text{AG}}$ (lowest frequency value). 
The vibrational spectrum extends from $\nu_{\text{AG}}$ up to $\sim$1600 cm$^{-1}$, although in the case for C$_{20}$, it is notorious a smaller frequency distribution range which goes from $\sim$480 cm$^{-1}$ to 1340 cm$^{-1}$.
It is well known that the C$_{20}$ fullerene is a controversial molecule, in which the reported calculated structure strongly depends on the utilized level of theory due to its high electron correlation and multiconfiguration character.~\cite{C20iso2002,C20exp2006,C20Hubbard2007,C20jt2009,C20ccsdt2015,SchNetNIPS2017,SchNet2018}
Often, different methodologies differ in the molecular point group of the cage ground state of C$_{20}$, which is reflected on the values of the interatomic forces and consequently in its vibrational spectrum. In particular, the main differences are located near the lowest frequency region.
In a previous theoretical work, Saito and Miyamoto~\cite{C20iso2002} reported a frequency distribution range that goes from $\sim$115 to $\sim$1435 cm$^{-1}$ using density functional theory (DFT) with the hybrid functional B3PW91. Another study by Sch\"utt \textit{el al.} recently reported a vibrational spectrum ranging from $\sim$100 to $\sim$1400 cm$^{-1}$ using a machine learning method based on deep neural networks, SchNet,~\cite{SchNetNIPS2017,SchNet2018} trained on DFT(PBE+vdW$^{\text{TS}}$)~\cite{TS} level of theory.
Similar results can be obtained with other machine learning approaches.~\cite{gdml,sgdml,sGDMLsoftware2019}
In these two works, the $\nu_{AG}$ are lower than in our case because the lower symmetry group considered as a global minimum, $D_{3h}$ instead of $I_h$. 
In the case of the highest frequency, also known as cut-off frequency $\nu_{\text{COF}}$, the lower value displayed in C$_{20}$ compared to other fullerenes is attributed to the nature of the carbon--carbon bond in C$_{20}$ coming from the fact that the geometrical structure is only formed by pentagonal faces.   
The rest of the studied fullerenes are more electronically stable and different level of theories give consistent results, showing a smoother transition between different sizes.
%
%

%
\begin{figure}[ht]
\centering
\includegraphics[width=1.0\columnwidth]{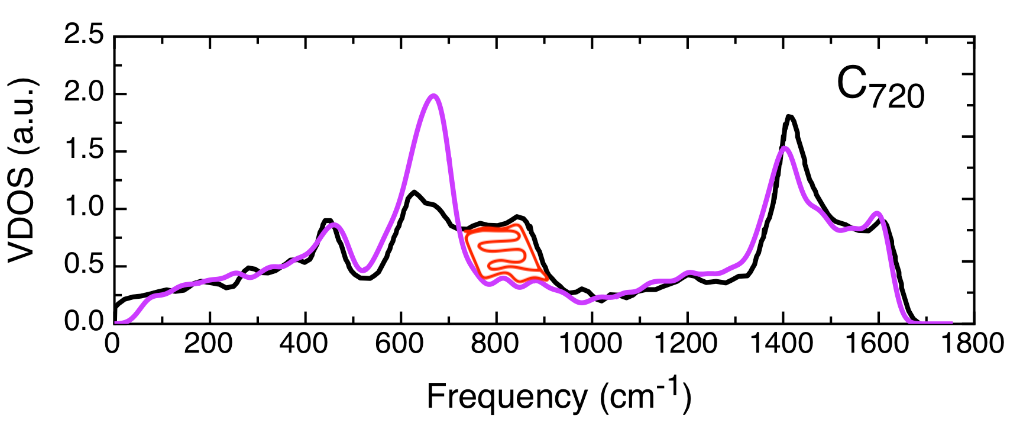}
\caption{VDOS comparison between fullerene C$_{720}$ (purple) and graphene (black). Bulk graphene VDOS was taken from Ref.\cite{Graphene2007} The suppressed phonon ZO branch in fullerenes is highlighted in red.}
\label{fig:Fig_C720Gr}
\end{figure}
\begin{figure*}[ht]
\centering
\includegraphics[width=1.0\textwidth]{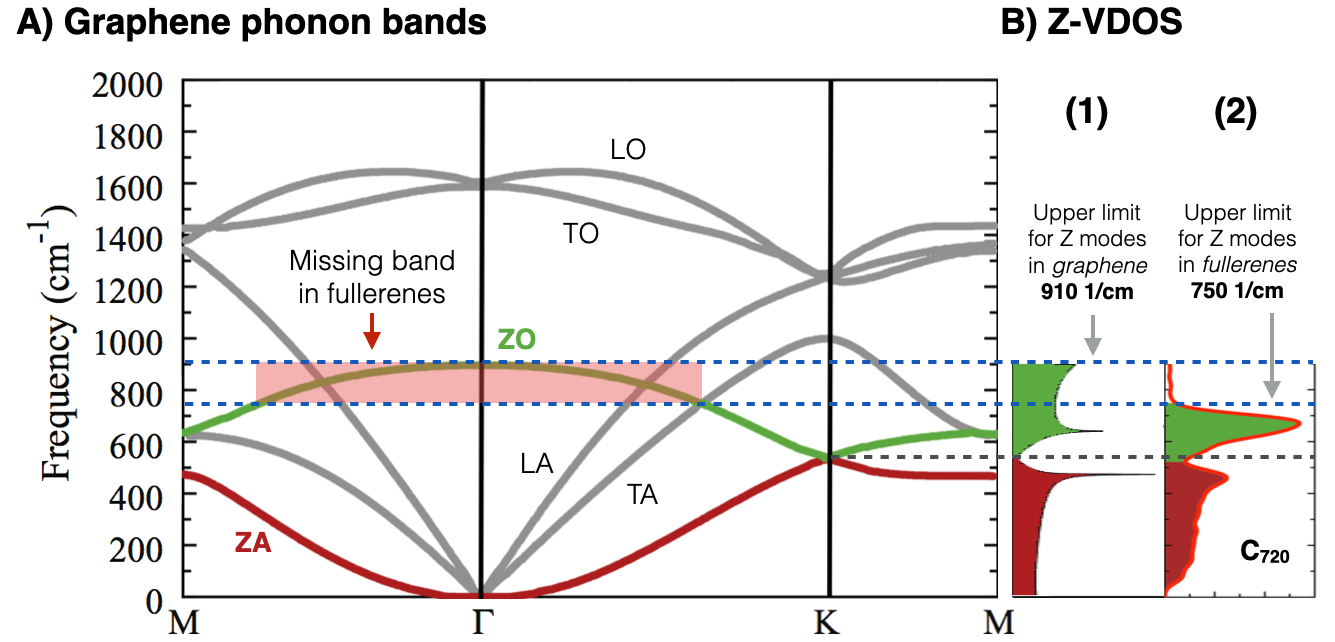}
\caption{Comparison of Z-phonon-bands between bulk graphene and fullerenes. A) The out of plane (Z) phonons are highlighted in red for acoustic (ZA) and green for optical (ZO) ones. B-1) ZO and ZA contribution to the graphene total VDOS. B-2) Radial (Z)-VDOS in C$_{720}$ fullerene. Graphene's VDOS was taken from Ref.~\cite{VDOSgraphene2013}}
\label{fig:Fig_PhononBands}
\end{figure*}
\begin{figure}[ht]
\centering
\includegraphics[width=1.0\columnwidth]{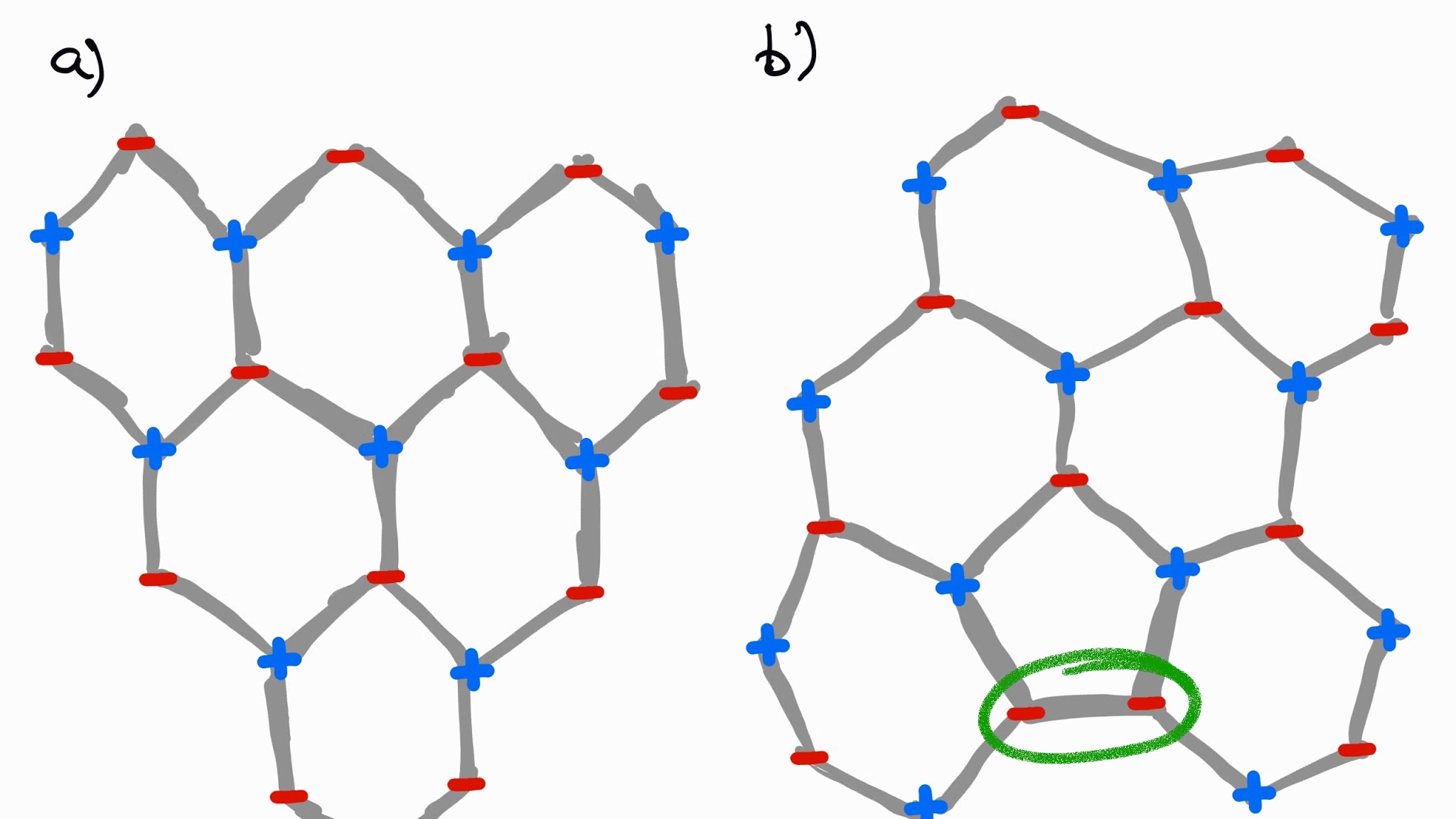}
\caption{a) Highest frequency ZO vibrational mode~\cite{VDOSgraphene2017} in graphene. b) Construction of a ZO vibrational mode in an hexagonal lattice with a pentagonal defect. In b) we can see that the presence of pentagons will produce an in-phase motion between first neighbors (marked in green), and consequently, the highest frequency ZO vibrational mode for fullerenes will be lower than the one in graphene.}
\label{fig:Fig_forbidden_ZO}
\end{figure}
%

\begin{figure*}[ht]
\includegraphics[width=1.0\textwidth]{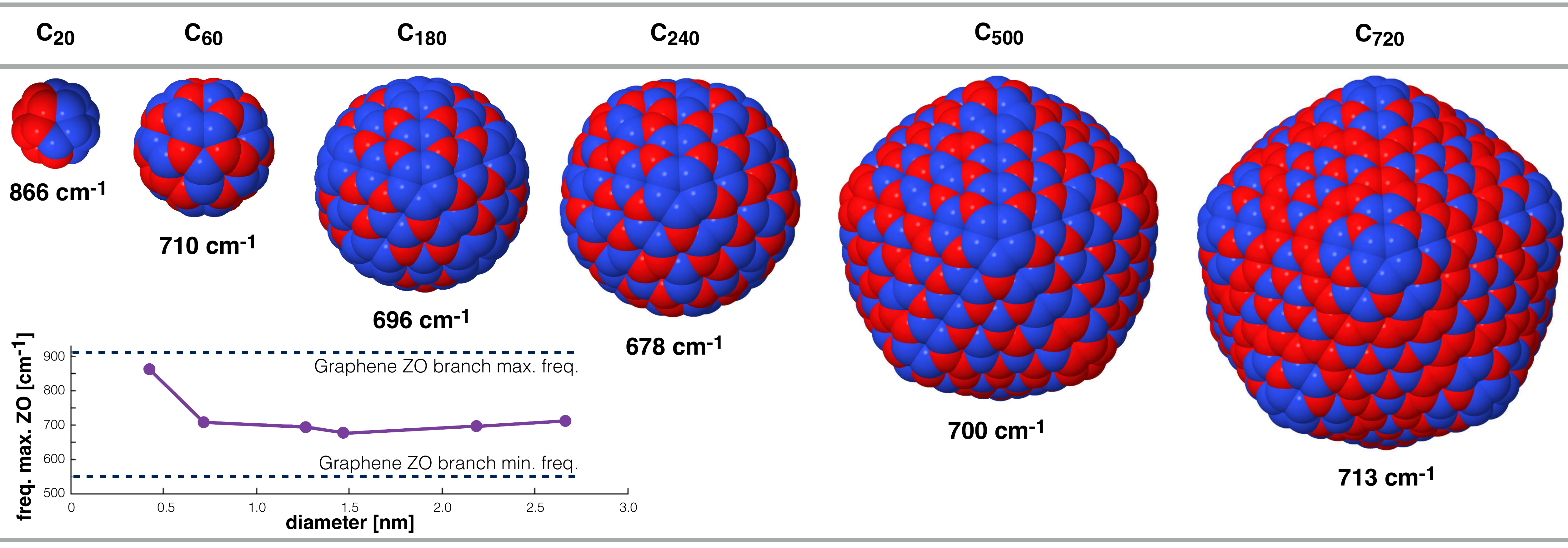}
\caption{Graphical representation of the highest frequency vibrational mode corresponding to purely radial motions, which the corresponds to the phonons in the ZO branch in graphene. The red and blue colors indicate phase and anti-phase radial atomic displacements, respectively. The inset plot shows the corresponding vibrational frequencies versus the fullerenes diameter. The bottom and top doted lines indicate the frequency limits of the ZO branch in graphene.}
\label{fig:Fig_ZOmodes}
\end{figure*}

\subsubsection{The hypothesis}

Based on the fact that fullerenes are single layer carbon cage structures, it is natural to assume that in the limit of large fullerene diameters, the VDOS should converge to the graphene's VDOS given its single layer character. 
To explore this hypothesis, in Fig.~\ref{fig:Fig_VDOSsize} we show the VDOS evolution with the size of the fullerenes and its comparison to
bulk graphene (Fig.~\ref{fig:Fig_VDOSsize} bottom panel).
Here we can think of the graphene case as the case of fullerene with diameter~$\to \infty$, where the twelve pentagonal faces are completely diluted.
%
From intermediate sizes, such as the C$_{240}$, we can see that the lineshape of the VDOS already has the main features of the biggest fullerene in our study (i.e. C$_{720}$). 
Furthermore, the fullerenes display the higher intensity (main) peak in the VDOS around 670 cm$^{-1}$, a characteristic feature that emerges from the 60 atom structure.

\subsubsection{Types of vibrational modes}

When comparing the VDOSs of the fullerene family and graphene, it is observed that the C$_{720}$ fullerene presents most of the features displayed by bulk graphene, with the obvious difference of the finite acoustic gap (See lowest two panels in Fig.~\ref{fig:Fig_VDOSsize}). 
However, it is worth to highlight the mismatch located in the range of $\sim$540--910 cm$^{-1}$, where the graphene peak around $\sim$850 cm$^{-1}$ is missing in the fullerene family (Fig.~\ref{fig:Fig_VDOSsize}-bottom and Fig.~\ref{fig:Fig_C720Gr}).
In order to understand this difference, first we analyze the VDOS of graphene. 
Graphene is a 2D material which presents vibrations/phonons that can be classified in two main groups, in-plane (XY) and out-of-plane (Z) vibrations (See Fig.~\ref{fig:Fig_PhononBands}). 
This means that in XY phonons the atomic displacement will be contained in the same plane of the graphene, whereas in Z phonons the displacements will be perpendicular to the graphene plane. 
Such decomposition of the VDOS for graphene has been discussed by  Paulatto et al.,~\cite{VDOSgraphene2013} indicating that the Z vibrational modes can be divided in acoustic (ZA) and optic (ZO) vibrations, and that their respective frequency intervals are [0, $\sim$540] and [$\sim$540, $\sim$910] cm$^{-1}$.~\cite{VDOSgraphene2013,VDOSgraphene2014,VDOSgraphene2017}
These two bands are highlighted in Fig.~\ref{fig:Fig_PhononBands}, ZA in red and ZO in green, as well as their own density of states, VDOS$^{ZA}$ and VDOS$^{ZO}$.
%
It is worth to stress that all Z vibrations/phonons are confined to frequencies \textit{below} 910 cm$^{-1}$.

By using this idea in the case of fullerenes, we can separate their VDOS contributions from normal modes with radial atomic displacements or out-of-shell (Z, red lines in Fig.~\ref{fig:Fig_VDOSsize}) from those containing displacements in-shell (XY, blue lines in Fig.~\ref{fig:Fig_VDOSsize}). 
In the case of the XY-VDOS for fullerenes (blue line), we observe a smooth line-shape spreading along the whole frequency spectrum, since it contains all the in-shell vibrational modes: TA, LA, TO and LO.
On the other hand, for the Z modes (red line) shown in Fig.~\ref{fig:Fig_VDOSsize} from top to bottom, we can see that the Z-VDOS converge to a smooth flat curve above 800 cm$^{-1}$, displaying two well defined peaks at 460 and 670 cm$^{-1}$, consistent with those observed in graphene (see Fig.~\ref{fig:Fig_VDOSsize}).

\subsection{The forbidden wave numbers in ZO branch in fullerenes}

As stated in the previous section, the Z modes in fullerenes have a cut-off frequency of $\sim$750 cm$^{-1}$, while the corresponding value for graphene is $\sim$910 cm$^{-1}$ (see Fig.~\ref{fig:Fig_PhononBands}).
Fig.~\ref{fig:Fig_C720Gr} shows the overlap between the VDOS of C$_{720}$ and graphene, clearly showing the fullerene's missing region in red.
By analyzing the phonon bands of graphene, we hypothesize that the missing peak is due to the lack of modes that would come from the highest frequency region of the ZO branch (centered in the $\Gamma$ symmetry point and marked by a red square in Fig.~\ref{fig:Fig_PhononBands}-A).
The origin of this is the fact that some symmetries of the wave vectors (i.e. vibrational eigenvectors) are not allowed in fullerenes.
To illustrate the kind of vibrational modes in the ZO (out-of-plane optical) branch, in Fig.~\ref{fig:Fig_forbidden_ZO}-a we show the highest frequency mode in the branch which corresponds to the high-symmetry $\Gamma$ point.
This consists in anti-phase out-of-plane atomic displacements between first neighbors.
Such vibrational mode is only allowed because of the hexagonal lattice of graphene (i.e. it has an even number of atoms in the ring), while in the case of fullerene family the presence of the pentagons prohibit the existence of this kind of vibrational mode. 
An schematic example is constructed in Fig.~\ref{fig:Fig_forbidden_ZO}-b, where the mode in Fig.~\ref{fig:Fig_forbidden_ZO}-a is being constructed from top to bottom in a lattice with a pentagonal ring, which then inevitably ends up in two neighbouring atoms moving in phase.
Therefore, \textit{the incapability of lattices with pentagons to host certain types of ZO modes limits the value of the cut-off frequency of Z-modes $\nu_{Z}^{max}$ in fullerenes.}
In fact, the frequency value $\nu_{Z}^{max}$ of the vibrational mode and its shape (i.e. its eigenvector) not only depend on the size but also on the molecular symmetry point group (see Fig. SI-1).
%
%
In general, the 12 pentagonal faces in fullerenes severely restrict high-frequency Z optical modes, but allows the creation of slower optical modes (hence the amplification of the ZO peak in Fig.~\ref{fig:Fig_VDOSsize} and~\ref{fig:Fig_PhononBands}-B).

%
%
%
This should also be true for bulk material with hexagonal lattice with pentagonal defects, since the odd number of atoms in the pentagonal ring hinders such highly symmetric vibrational motion, originating a lower frequency in-phase motion between first neighbors (Fig.~\ref{fig:Fig_forbidden_ZO}-b, in green).

Figure ~\ref{fig:Fig_PhononBands} highlights the  contributions of the ZO (green) and ZA (red) phonon bands to the total VDOS, and their direct comparison to the fullerences' Z-VDOS is presented.
From this figure we can clearly see the region of the ZO phonon branch that would correspond to the missing vibrational modes in the fullerene family.

\subsection{Visualization of highest Z modes}
In order to visually understand what was explained above, it is interesting to analyze how the ZO modes look like in fullerenes, in particular the one corresponding to the maximum allowed frequency, $\nu_Z^{max}$.
The family of these modes is displayed in Fig.~\ref{fig:Fig_ZOmodes}. 
From here, we can see that starting from C$_{180}$ to C$_{720}$, the shape of the associated eigenvector $\textbf{v}_Z^{max}$ follows similar patterns.
The shape of the vibrational modes can be contrasted to the highest frequency ZO phonon ($\Gamma$ point) in Fig.~\ref{fig:Fig_forbidden_ZO}-a. 
Interestingly, the $\nu_Z^{max}$ value converges to $\sim$710 cm$^{-1}$ for all fullerenes in our study, with the exception of C$_{20}$ given the fundamental differences in the structure and chemical bonds.
The inset plot in Fig.~\ref{fig:Fig_ZOmodes} shows the range of frequencies in which the ZO branch is defined in graphene (highlighted in green in Fig.~\ref{fig:Fig_PhononBands}) and where the $\nu_Z^{max}$ values lay relative to this reference interval.
It is  expected that as the diameter of the fullerene grows, the $\nu_Z^{max}$ will slowly start increasing towards the graphene ZO branch maximum value given that the contribution of the area of the 12 pentagons relative to the rest of the fullerene surface will start diluting.

\section{Discussion}
The present analysis  on the evolution of fullerenes' VDOS with the diameter and its similarities and deviations from the bulk graphene VDOS reveals an interesting geometrical constraint on the type of vibrational modes that fullerenes can host.
Unlike the size evolution towards the bulk behavior obtained for the VDOS of metal nanoparticles, \cite{Sauceda_VDOS_2015a} in the present case there will not be a full convergency to the graphene VDOS since the pentagonal facets are required to close the cage structures of fullerenes. 
In particular, in terms of the phononic bands in graphene, the ZO-equivalent normal mode oscillations are greatly affected by the presence of pentagonal faces, which sets a lower frequency ($\sim$750 cm$^{-1}$) and dynamically different ZO cut-off eigenstate (comparison between Fig.~\ref{fig:Fig_forbidden_ZO}-a and Fig.~\ref{fig:Fig_ZOmodes}), as well as a narrower ZO band ($\sim$43\%).
This compression of the ZO-band in fullerenes, reshapes the VDOS, as shown in Fig.~\ref{fig:Fig_C720Gr}, which means that necessarily this will have implications on the thermodynamics of the system.
In particular, it is expected that this would induce a distinct behavior in the thermal properties at low temperatures of fullerenes and graphene.

An important assumption in our study was that we can take graphene's VDOS as a fullerene with a very large diameter, where the effects of the pentagonal faces is not relevant.
Hence, this implies that $\lim_{d\to\infty}{\nu_{Z}^{max}(d)}=\nu_{ZO,Graphene}^{max}$, where $d$ is the diameter of the fullerene.
Such convergence is very slow as can be appreciated in Fig.~\ref{fig:Fig_ZOmodes}-inset.
Nevertheless, we have to remember that normal modes are collective atomic oscillations, meaning that either the interatomic interaction in the pentagonal faces strengthen with the size (thereby increasing its frequency) or the pentagonal faces slowly become nodes in the normal mode oscillation i.e. in the $\textbf{v}_Z^{max}$ eigenvector, allowing then higher frequency oscillations localized in the hexagons.
By analysing the atomic amplitudes of the normal modes in Fig.~\ref{fig:Fig_ZOmodes}, we found evidence of the latter option. 
Such result would imply that, if the vibrational mode corresponding to the $\Gamma$ symmetric point in the ZO phonon band exists in a graphene material with pentagonal defects, those defects will have null amplitudes.
A thorough analysis in this topic requires further evidence through numerical calculations and goes beyond the scope of this article, then will be left as future work.

To conclude, we have presented here a thorough analysis of the peculiarities in the vibrational properties of fullerences relative to the bulk graphene, and their potential thermodynamical differences due to the presence of a compressed \textit{ZO-band} in fullerenes.
This work should open question regarding the implications and dynamics of pentagonal or other type of ring defects on graphene, but also it fills up a  knowledge-gap in our understanding of the physical properties of carbon nanostructures.

\section{METHODOLOGY}
 All computations were done at the DFT level within the generalized  gradient approximation (GGA), using the Perdew-Burke-Ernzerhof (PBE) parameterization.~\cite{PBE1996} A split valence basis set (def2-SVP) and the pseudopotential for carbon with 4 valence electrons corresponding to ECP2SDF were used,~\cite{pseudopotentials,ScalingLawVDW} as implemented in the TURBOMOLE code.~\cite{TURBOMOLE}
To guarantee accurate frequencies calculation, the structures were optimized with a force tolerance of 10$^{-6}$ Hartree Bohr$^{-1}$. 
%
The computed harmonic frequencies were validated with reported calculations for carbon dimer's bond length, binding energy, and vibrational frequency.~\cite{hbcp,Blanksby,Irikura,C60exp1991,C60IR1991,C60freqs2000,C60freqs2001}
Additionally, the whole vibrational spectrum of C$_{60}$ fullerene was calculated and compared with experimental results~\cite{RamanIRc60} for validation. 
The VDOS was constructed using a Gaussian broadening with a width of 20 cm$^{-1}$ of the $3N-6$ frequencies. 

The selection of purely radial vibrational modes in fullerenes and their contribution to the VDOS was estimated by performing a radial projection of the whole vibrational spectrum, $\text{Z-VDOS}(\nu)=\sum_{i=1}^{3N-6}\langle \textbf{x}|\textbf{v}_i\rangle \delta(\nu-\nu_i)$, where $\textbf{x}$ is a radial vector and $\{\textbf{v}_i\}$ is the set of eigenvectors. 
Then, the in-plane (or more precisely for fullerenes, in-shell) contribution is computed via XY-VDOS=VDOS -- Z-VDOS.


\section{Acknowledgements}
ILG thanks support from DGTIC-UNAM under Project LANCAD-UNAM-
DGTIC-049, DGAPA-UNAM uder Project IN106021, and CONACYT-Mexico under Project 285821.




\bibliography{references}

\end{document}